\documentstyle[emulateapj,apjfonts,psfig]{article}

\submitted{ApJ accepted version June 6, 2006}
\lefthead{J.\ E.\ Geach et al.}
\righthead{Spitzer MIPS surveys of two distant clusters}

\def\gs{\mathrel{\raise0.35ex\hbox{$\scriptstyle >$}\kern-0.6em
\lower0.40ex\hbox{{$\scriptstyle \sim$}}}}
\def\ls{\mathrel{\raise0.35ex\hbox{$\scriptstyle <$}\kern-0.6em
\lower0.40ex\hbox{{$\scriptstyle \sim$}}}}

\newenvironment{inlinefigure}{%
\def\@captype{figure}%
\noindent\begin{minipage}{0.999\linewidth}\begin{center}}
{\end{center}\end{minipage}\smallskip}
\makeatother

\begin{document}

\title{A panoramic mid-infrared survey of two distant clusters}

\author{J.\ E.\ Geach\altaffilmark{1}, Ian\ Smail\altaffilmark{2}, R.\ S.\ Ellis\altaffilmark{3}, S.\ M.\ Moran\altaffilmark{3}, G.\ P.\ Smith\altaffilmark{3,4}, T.\ Treu\altaffilmark{5}, J.-P.\ Kneib\altaffilmark{6},\\ A.\ C.\ Edge\altaffilmark{2} \& T.\ Kodama\altaffilmark{7}}

 \altaffiltext{1}{Department of Physics, Durham University, South Road, Durham DH1\ 3LE. UK.\ j.e.geach@durham.ac.uk}

 \altaffiltext{2}{Institute for Computational Cosmology, Department of Physics, Durham University, South Road, Durham DH1\ 3LE. UK.}

 \altaffiltext{3}{Department of Astronomy, California Institute of Technology, MC\ 105-24, Pasadena, CA\ 91125}

 \altaffiltext{4}{School of Physics and Astronomy, University of Birmingham, Edgbaston, Birmingham, B15\ 2TT.\ UK} 

 \altaffiltext{5}{Physics Department, University of California, Broida Hall, Mail Code 9530, Santa Barbara, CA 93106--9530}

 \altaffiltext{6}{OAMP, Laboratoire d'Astrophysique de Marseille, UMR 6110 traverse du Siphon, 13012 Marseille. France}

 \altaffiltext{7}{Optical and Infrared Astronomy Division, National Astronomical Observatory, Mitaka, Tokyo 181-8588. Japan}

\begin{abstract}
We present panoramic {\it Spitzer MIPS} 24-$\mu$m observations covering
$\sim$9$\times$9\,Mpc ($25'\times 25'$) fields around two massive
clusters, Cl\,0024+16 and MS\,0451$-$03, at $z=0.39$ and $z=0.55$
respectively, reaching a 5-$\sigma$ flux limit of $\sim 200\mu$Jy. Our
observations cover a very wide range of environments within these
clusters, from high-density regions around the cores out to the
turn-around radius.  Cross-correlating the mid-infrared catalogs with
deep optical and near-infrared imaging of these fields, we investigate
the optical/near-infrared colors of the mid-infrared sources.  We find
excesses of mid-infrared sources with optical/near-infrared colors
expected of cluster members in the two clusters and test this selection using spectroscopically confirmed 24$\mu$m members.  The much more
significant excess is associated with Cl\,0024+16, whereas
MS\,0451$-$03 has comparatively few mid-infrared sources.  The
mid-infrared galaxy population in Cl\,0024+16 appears to be associated
with dusty star-forming galaxies (typically redder than the general
cluster population by up to $A_V\sim1$--2\,mags) rather than emission
from dusty tori around active galactic nuclei (AGN) in early-type
hosts.  We compare the star-formation rates derived from the total
infrared (8--1000$\mu$m) luminosities for the mid-infrared sources in
Cl\,0024+16 with those estimated from a published H$\alpha$ survey,
finding rates $\gs5\times$ than those found from H$\alpha$,
indicating significant obscured activity in the cluster population.
Compared to previous mid-infrared surveys of clusters from
$z\sim0$--0.5, we find evidence for strong evolution of the level
of dust-obscured star-formation in
dense environments to $z=0.5$, analogous to the rise
in fraction of optically-selected star-forming galaxies seen in
clusters and the field out to similar redshifts. However, there are
clearly significant cluster-to-cluster variations in the populations of
mid-infrared sources, probably reflecting differences in the
intracluster media and recent dynamical evolution of these systems.
\end{abstract}

\keywords{galaxies: clusters -- galaxies: starburst -- galaxies:
  infrared -- clusters: individual Cl\,0024+16, MS\,0451$-$03}

\section{Introduction}

\begin{figure*}
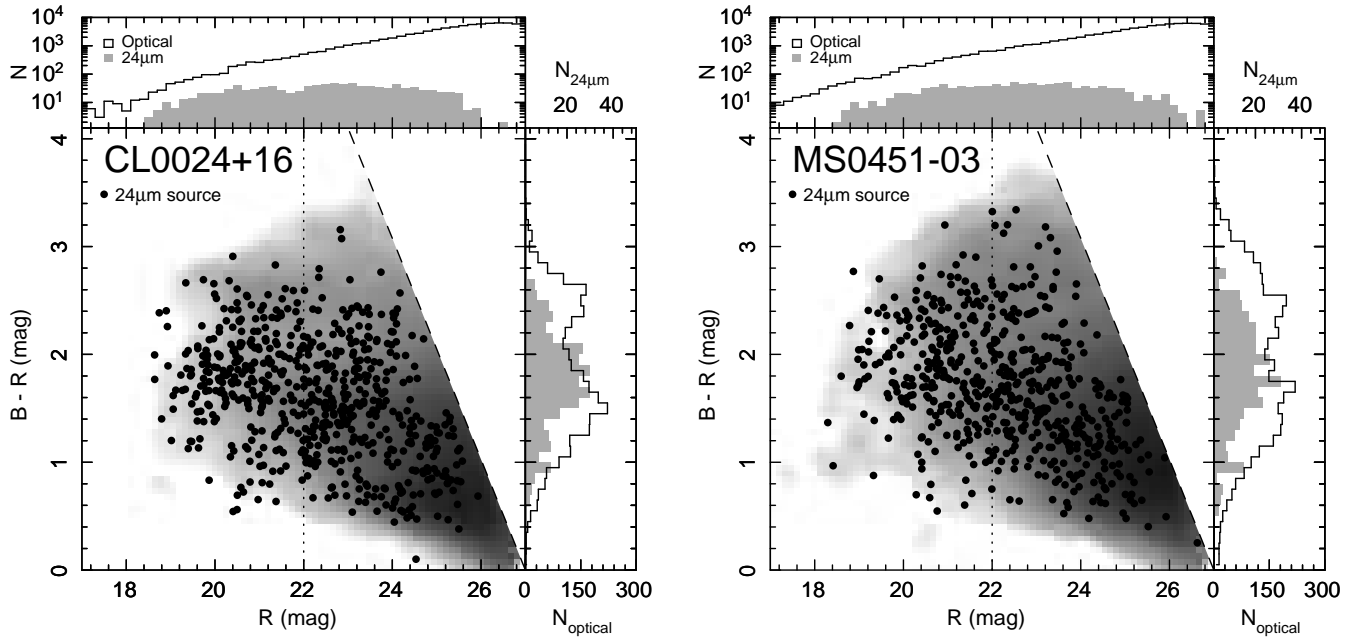

\centerline{
\psfig{file=f1a.ps,width=3.5in,angle=0}~~
\psfig{file=f1b.ps,width=3.5in,angle=0}}

\caption{\small The $(B - R)$--$R$ color-magnitude diagram for (left)
Cl\,0024+16 and (right) MS\,0451$-$03. We identify 24-$\mu$m sources
brighter than 200$\mu$Jy and compare these to the distribution for the
optically-selected populations in these fields (for clarity, this has
been represented by a smoothed density plot).  Also shown are
histograms showing slices through the distributions, comparing the
number counts and colors for the mid-infrared sources to the
optically-selected population.  To emphasise the location of the
sequence of early-type galaxies in the clusters, the $(B-R)$ histogram
is limited to $R\leq 22$. The mid-infrared selected population lies
between the red and blue galaxy peaks in these fields, most likely
because of the influence of dust on intrinsically blue, star-forming
galaxies.  For the mid-infrared sources in the color-magnitude diagram
we have removed those objects morphologically classified as stars.  All
colors are measured in 2$''$ diameter apertures, and we use the {\sc
best\_mag} estimate of the total galaxy magnitude. }
\end{figure*}

\begin{figure*}[t]
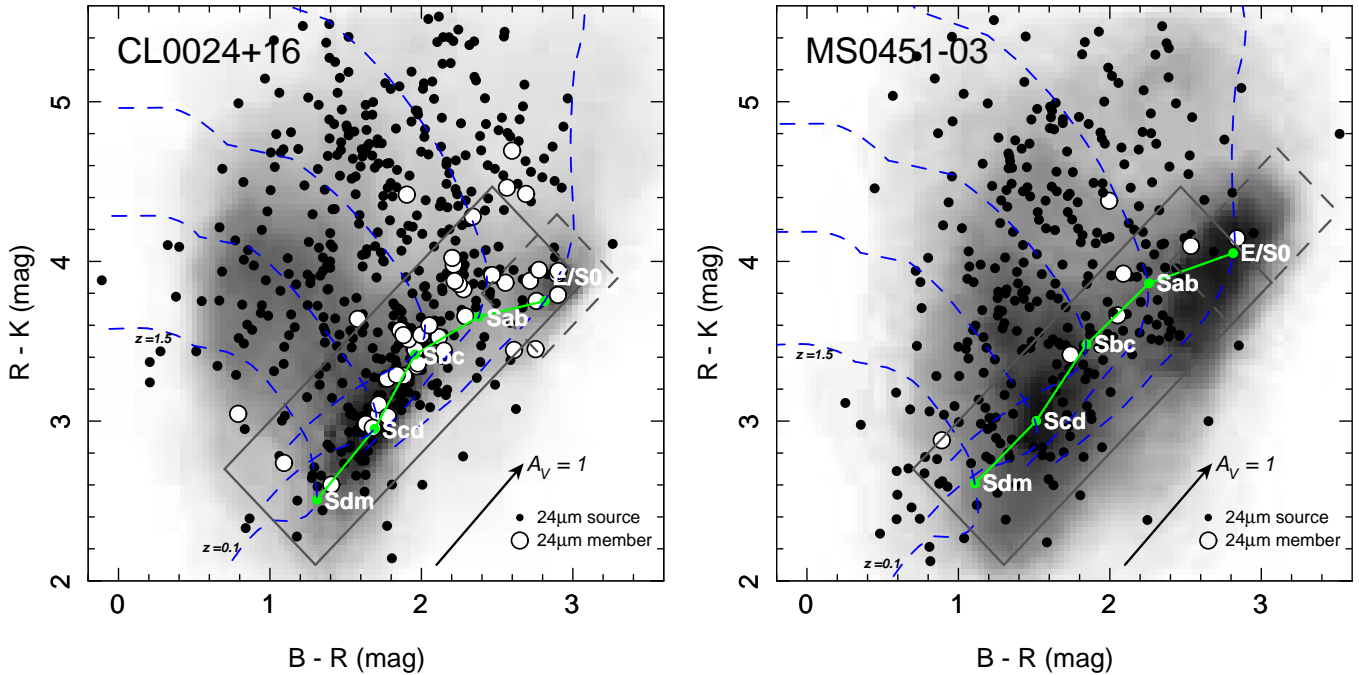

\centerline{
\psfig{file=f2a.ps,width=3.5in,angle=0}~~
\psfig{file=f2b.ps,width=3.5in,angle=0}}
\caption{\small We show $(R-K)$ versus $(B-R)$ colors for
24$\mu$m-detected sources compared to the optical distribution in the
Cl\,0024+16 (left) and MS\,0451$-$03 (right) fields.  The distributions
are limited at $R\leq24$ and $K\leq20$ and we represent the optical
sample's number density as a slightly smoothed grayscale.  The dashed
tracks trace the expected colors of galaxies across $z=0$--2, with star
formation histories similar to that of present-day Sdm--E galaxies, and
the solid line and points show the location in color--color space of
the range of spectral types at the cluster redshift.  The arrow
indicates the translation in color--color space corresponding to an
increase in reddening of $A_V=1$. To perform a rough photometric cut to
select 24$\mu$m sources at the cluster redshift, we define a region
around the predicted colors of cluster members indicated by the
rectangular selection boxes in both panels. To avoid contamination by infrared
emission from AGN, we designate a sub-region which should contain the
passive galaxies (dashed-line box) -- 24$\mu$m sources in this region
are not included in our analysis of the obscured star-forming
populations. We show the efficacy of this selection by plotting the
colors of known spectroscopically-confirmed 24$\mu$m cluster members,
85\% of these fall within the selection box.}
\setcounter{figure}{2}
\end{figure*}

The galaxy populations within the virialised regions of rich clusters
at $z\sim 0$ are characterised by passive elliptical and lenticular
(S0) galaxies (Oemler 1974; Dressler 1980).  In contrast, 5-Gyrs ago,
at $z\sim 0.5$, the galaxy populations in the most massive clusters
had larger fractions of star-forming late-type spirals, and a
corresponding deficit of luminous S0 galaxies (e.g.\ Dressler et
al.\ 1997; Smail et al.\ 1997; Couch et al.\ 1998; Fasano et al.\ 2000;
Treu et al.\ 2003).  Taken together, these two observations imply that
a process (or processes) is transforming many of the star-forming,
late-type spirals in these regions into the passive early-type
population (specifically S0s) found in local clusters (Poggianti et
al.\ 1999; Moran et al.\ 2006, in prep).

When considering potential pathways to produce this evolutionary
change, we need to bear in mind that the typical luminosities of the
star-forming spirals appear to be too low for them to transform into
typical S0 galaxies found in local clusters, without the addition of
significant numbers of new stars (Poggianti et al.\ 1999; Kodama \&
Smail 2001).  This problem is exacerbated when we include the fading
which is likely to take place after the cessation of star formation in
these galaxies.  This then leads us to concentrate on mechanisms which
are capable of increasing the luminosity of the galaxies -- mergers and
starbursts.  There has been a recent upsurge in interest in the
potential for so-called ``dry'' mergers (mergers between
dissipationless stellar systems which don't result in additional star
formation) to influence the evolution of early-type galaxies (van
Dokkum et al.\ 2003; Bell et al.\ 2003; van Dokkum 2005).  However, the
dynamically hot environments in rich clusters which are the subject
of our study are deleterious to the formation and survival of cold,
bound-pairs of early-type galaxies -- unless these systems arrive in
the cluster as existing bound entities.  It is not clear, therefore,
that dry mergers can provide an effective route to substantially
increase the number of luminous, early-type S0 galaxies within clusters.

Unfortunately, the alternative mechanism for enhancing the luminosity
of the bulge component -- a starburst -- also has strong observational
evidence stacked against it.  Surveys of star-forming galaxies in
clusters using optical or UV star formation indicators have failed to
detect galaxies with strongly enhanced star-formation which would have
to exist to explain the growth of the bulge components of early-type
galaxies in clusters at $z\ls 0.5$--1 (Balogh et al.\ 1999; Poggianti
et al.\ 1999; Gerken et al.\ 2004).  However, there is growing body of
evidence that at least some of the galaxies in distant clusters may be
undergoing bursts of star-formation, albeit ones which are heavily
shrouded in dust.  Smail et al.\ (1999) used a deep VLA 1.4-GHz radio
map to study a small sample of active galaxies within the core of the
cluster Cl\,0939+4713 ($z=0.41$). Combining the radio data with
near-infrared and optical morphological information from the {\it
Hubble Space Telescope (HST)} and ground-based spectroscopy, they found
that over half the radio-emitting population in the core are dusty 
late-type galaxies, presumably undergoing vigorous star formation. However,
the spectral classification of these spirals placed them in the
post-starburst class, and indeed all the post-starburst galaxies in this small
region are radio-emitters.  Dust has also been used to explain the
unusual spectral properties of another class of galaxies found in
distant clusters and the field: e(a) galaxies, which show enhanced
Balmer absorption compared to normal star-forming galaxies (Poggianti
et al.\ 1999).  Poggianti \& Wu (2000) and Poggianti, Bressan \&
Franceschini (2001) discuss models for these galaxies invoking
age-dependent dust obscuration of the younger stellar populations --
enabling significant activity to be hidden from view in these
systems. If the passive lenticular galaxies found in local clusters,
but absent from the equivalent rich environments at higher redshift, are
the result of infalling late-type galaxies undergoing dusty-starburst
in high-$z$ clusters, then a possible signature would be evolution in
the total level of obscured star-formation in clusters out to $z\sim1$.
 
In principle, mid- and especially far-infrared/submillimeter
observations give us a direct probe of the level of obscured activity
in distant clusters.  In particular, mid-infrared observations with
{\it Infrared Space Observatory (ISO)}, and more recently {\it Spitzer
Space Telescope (SST)}, provide sensitive imaging capabilities which can
trace dusty star formation in clusters out to $z\sim 1$ and beyond.
Metcalfe, Fadda \& Biviano (2005) summarise the results from {\it ISO}
surveys of distant clusters, which have yielded a total of just $\sim
40$ cluster galaxies detected at 15\,$\mu$m across seven clusters
between $z=0.18$--0.56 (Duc et al.\ 2000, 2004; Fadda et al.\ 2000;
Metcalfe et al.\ 2003; Coia et al.\ 2005a,b; Biviano et al.\ 2004).
All these studies suggest that there is an increased level of
mid-infrared activity in distant clusters, at levels above that
suggested by UV/optical tracers of star formation.  Submillimeter
observations of more distant clusters have also hinted at possible
enhanced activity in these environments (Best 2000; Webb et al.\ 2005).
However, the inhomogeneous mix of coverage and depth in the samples
coupled with the modest numbers of sources detected in any individual
cluster mean that it has proved difficult to use these data to provide
quantitative constraints on the origin and evolution of dust-obscured
activity in distant clusters.

The {\it SST}'s sensitive mid-infrared imaging capabilities provide an
unique opportunity to undertake complete and representative surveys of
the obscured, active populations in distant clusters.  To search for a
population of mid-infrared sources in rich clusters environments, we
have therefore used the Multiband Imaging Photometer for {\it Spitzer}
(MIPS) to detect 24-$\mu$m emission from galaxies in two clusters at
$z\sim0.5$ covering a very wide range in environment from $\sim1$\,Mpc
out to the turn-around radius ($\sim5$\,Mpc) where the clusters merge
into the surrounding field. These observations will provide measures of
the level of obscured star-formation in these clusters, and so allow us
to build up a reliable picture of the evolution of dust-obscured
activity in clusters over the past 5\,Gyrs.

This paper presents a statistical analysis of the 24$\mu$m populations
in two $z\sim 0.5$ clusters. A subsequent paper (Geach et al.\ in prep)
will discuss the properties of these sources in more detail using the
available spectroscopic and morphological surveys of the clusters
(Moran et al.\ 2006). The paper is organised as follows: we
describe our observations and their reduction in \S2, analyse these in
\S3 and discuss our results and present our conclusions in \S4 and \S5,
respectively.  Throughout, we adopt a geometry with $\Omega_m=0.3$,
$\Omega_\Lambda=0.7$ and $H_0 = 70$\,km\,s$^{-1}$\,Mpc$^{-1}$.

\section{Observations \& Reduction}

The two clusters chosen for this study are unique in having panoramic
{\it HST} imaging covering $\sim 25'$-diameter fields -- extending from
the cores out to the turn-around radii of the clusters (Treu et al.\
2003; Kneib et al.\ 2003; Moran et al.\ 2005).  These data have been
used for weak-lensing analysis of these clusters, yielding 2-D maps of
the dark matter distributions on $\sim 5$-Mpc scales in the structures
(Kneib et al.\ 2003).  Panoramic studies of the galaxy populations in
these clusters also benefit from extensive deep, ground-based optical
and near-infrared imaging and spectroscopy (Moran et al.\ 2006 in prep).  Although both clusters are relatively rich, they differ in their  X-ray luminosities: Cl\,0024+16 ($z = 0.39$) has a relatively
modest X-ray luminosity, $L_X\sim 3.2\times10^{44}$\,ergs\,s$^{-1}$
(Treu et al.\ 2003), while MS\,0451$-$03 ($z = 0.55$) is some 8$\times$
more luminous. This distinction may lead to differences in the
effectiveness of the various processes
influencing star formation (Treu et al.\ 2003), as traced by the
distribution of the mid-infrared population. This will be useful in our
subsequent detailed study to disentangle the potential mechanisms for
triggering and suppressing star formation.

\subsection{Mid-infrared observations}

MIPS 24$\mu$m observations of the fields of Cl\,0024+16 ($z=0.39$) and
MS\,0451$-$03 ($z=0.55$) were obtained with {\it SST} in fixed-cluster
offset mode on 2004 December 24--25.  The observations of Cl\,0024+16
are centered on 00\,26\,35.70, $+$17\,09\,45 (J2000); while those for
MS\,0451$-$03 are centered on: 04\,54\,10.80, $-$03\,00\,57 (J2000).
The observations avoided the central $\sim 5{'} \times 5{'}$ of each
cluster, which are part of Guaranteed Time Observations ({\it SST}
Program \#83). For MS\,0451$-$03 we have acquired these data from the
archive, and incorporated them into this work to provide coverage of
the central region of the cluster. The observations of the core of
Cl\,0024+16 are not available at the time of writing, however there are
existing mid-infrared observations from {\it ISOCAM} at 15$\mu$m (Coia
et al.\ 2005), which traces similar emission to that sampled by MIPS
(namely emission from polyaromatic hydrocarbons). These allow us
to place limits on the mid-infrared population in 
the central regions of Cl\,0024+16.

Each field was covered in a $5\times5$ grid (omitting the central
pointing) yielding a 24-point mosaic. We adopted two 10-s cycles for
our exposures producing a per-AOR pixel exposure of 312.5s, and the
total elapsed time for the AOR was 9.78\,ks. The template AOR was
repeated three times to provide redundancy and for the identification
of cosmic rays and asteroids (the latter can be common in 24$\mu$m MIPS
data). The total exposure time per pixel is 938\,s, which should
achieve our goal of a 5-$\sigma$ detection limit of 200$\mu$Jy at
24$\mu$m.

The basic-calibrated 24$\mu$m data provided by the {\it Spitzer}
Science Center were first corrected for gradients, bright latents and
the ``jailbar'' effect using calibration frames generated from the
data. These were then mosaiked using the {\sc mopex}
package.\footnote{{\sc mopex} is maintained by the {\it Spitzer}
Science Center: http://ssc.spitzer.caltech.edu} Due to the `first frame
effect' (frames with a shorter exposure and a depression in response of
up to 15\%), the initial frame in each pointing was discarded, which
improved flatness in the final mosaic. We implemented the the area-overlap interpolation method in the mosaicking procedure (Makovoz \& Khan\ 2005). The mosaics cover total areas of $\sim$\,0.21 sq.\ degrees in both clusters.

\begin{inlinefigure}
\centerline{\psfig{file=f3.ps,width=3.5in,angle=0}}
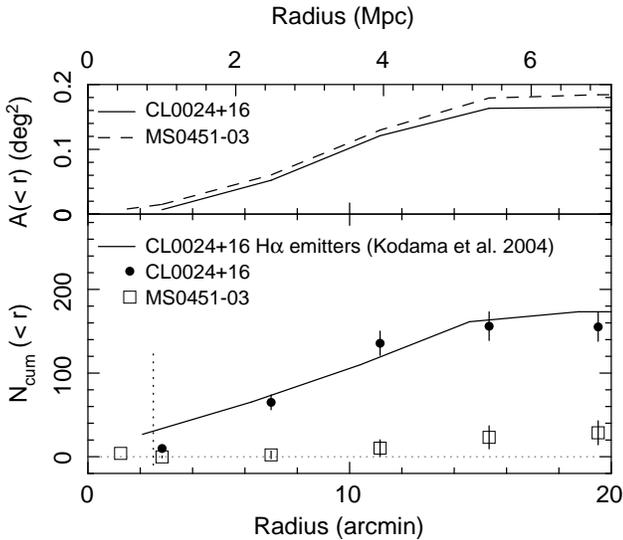
\begin{minipage}{3.5in}{{\small \sc Fig.~3.}--\small\ The lower panel shows the cumulative number of probable
24$\mu$m cluster members (brighter than 200$\mu$Jy) as a function of
radial separation from their respective cluster centers.  Probably cluster members are isolated based on their $(R-K)$--$(B-R)$ colors (see
Fig.~2). We correct the distributions for residual field contamination
using a similar
color-cut on the SWIRE ELAIS N1 survey (Lonsdale et al.\ 2003,\ 2004, Surace et al.\ 2004), giving an estimated field
correction of $\sim0.16\pm 0.02$ arcmin$^{-2}$.  The estimated numbers
of members could also be increased by $\sim 18$\% to account for
the incompleteness in our color-selection, although we have not done so here.
We also plot the cumulative
number of H$\alpha$-emitting (EW(H$\alpha$) $>$40\AA) cluster members in Cl\,0024+16 from the
survey by Kodama et al.\ (2004), which exhibits a similar rise to that
exhibited by the 24$\mu$m population in this cluster.  In contrast,
MS\,0451$-$03 shows little evidence for a strong mid-infrared
population.  
The top panel illustrates the variation
in survey area with radius in our two fields, this shows that the
slower increase in the cumulative 24$\mu$m counts beyond 5\,Mpc in
Cl\,0024+16 is purely due to the decline in coverage in the outskirts
of our survey. The errors on our observations are the combination of
counting statistics and the measured variance in the field number
counts. The physical scale assumes an average redshift of $z=0.47$.}\end{minipage}
\setcounter{figure}{3}
\end{inlinefigure}

Source extraction was performed using {\sc SExtractor} version 3.1
(Bertin \& Arnouts 1996) with the criteria that a source consists of at
least 3 contiguous pixels (each pixel is 2.5${''}$ square) at $\geq
2\sigma$ above the background.  We measure 16-${''}$ diameter aperture
fluxes, corresponding to $\sim3\times$ FWHM of the PSF (FWHM of $\sim
5''$ at 24$\mu$m). Using a curve-of-growth analysis on bright isolated
point sources, and our completeness simulations, we find that 16${''}$
apertures recover $\sim75$\% of the total flux and we therefore correct
the resulting fluxes by a factor of $1.33\times$ to yield total
fluxes. We visually inspect the extracted objects overplotted on the
images, and remove sources which appear to be false, e.g.\ in the
slightly noisier regions on the edge of the image.

The flux detection limits and completeness of each mosaic was
determined by inserting 10 sets of 10 simulated sources into the
mosaics for a range of flux and determining their detection rate and
recovered fluxes using identical extraction parameters.  The sources
used in the simulations were high signal-to-noise composites formed by
stacking a large number of real, isolated faint sources. From these
simulations we estimate that the 80\% completeness limits of our
observations are 180$\mu$Jy in Cl\,0024+16 and 200$\mu$Jy in
MS\,0451$-$03. 

The archival mid-infrared data for the central region of MS\,0451$-$03
(Program \#83), were obtained from the {\it Spitzer} archive. The
observations were taken on 2004 September 23, centered on 04\,54\,10.8,
$-$03\,00\,57 (J2000). We use the post--basic-calibrated data, and
subject it to the same extraction criteria as described in \S2.1. These
data are slightly shallower than our mosaic, and we take this into
account in our subsequent analysis.

\subsection{Archival optical \& near-infrared imaging}

We use existing deep optical and near-infrared imaging to obtain colors
for the 24$\mu$m sources in these regions. For both clusters we use
Subaru SuprimeCam $B$- and $R$-band observations taken for the PISCES
survey (Kodama et al.\ 2005). The Cl\,0024+16 field was observed on
2002 September 6 under good conditions. The seeing was $\sim0.7$--$1''$
for the $R$- and $\sim1$--$1.3''$ for the $B$-band, with exposure times
of 5,280\,s and 3,600\,s respectively, reaching a depth of $R\sim27$
(we use Vega-based magnitudes throughout). The observations and data
reduction technique are described in Kodama et al.\ (2004).
MS\,0451$-$03 was observed on 2001 January 21--22, again in good
conditions with 1.0$''$ seeing in the $B$-band and 0.8$''$ in the
$R$-band.  Total exposure times were 7200\,s in $B$ and 4800\,s in $R$,
again yielding photometry for objects as faint as $R\sim 27$.

Panoramic near-infrared $K$-band imaging of both clusters is also
available with WIRC (Wilson et al.\ 2003) on the Palomar Hale 5.1-m
telescope. These data comprise a $3\times3$ mosaic of WIRC pointings,
providing a contiguous observed are of $26'\times26'$ centered on each
cluster. Full details of the observations and data reduction are
published elsewhere (Cl\,0024+16: Smith et al.\ 2005; MS\,0451$-$03:
Smith et al.\ 2006, in prep). Point sources in the final reduced mosaics
have a full width half maximum of $0.9''$ and $1.0''$ in Cl\,0024+16
and MS\,0451$-$03 respectively. The data reach a 5-$\sigma$ point sources detection threshold of $K=19.5$
and $K=20.0$ in Cl\,0024+16 and MS\,0451$-$03 respectively. The nominal 24$\mu$m coverage of the two clusters is $\sim$0.21\,degrees$^2$. In the case of MS\,0451$-$03, the archival GTO data contribute an extra $\sim$29\,arcmin$^2$. When the optical and near-infrared coverage is taken into account, the total coincident coverage of 24$\mu$m, $B$-, $R$-, $K$-bands is 0.165\,degrees$^2$ and 0.184\,degrees$^2$ in Cl\,0024+16 and MS\,0451$-$03 respectively. 

We detect objects and extract photometry from the optical frames by
running {\sc SExtractor} in two-frame mode such that detections were
made in the $R$-band image and measurements made at identical pixel
locations in $R$ and $B$ (the frames were previously aligned to good
accuracy). We catalog all sources with 3 contiguous pixels (the pixel
scale is $0.204''$) at least 2-$\sigma$ above the background, and lay
down 2$''$ diameter apertures to measure colors in the $BRK$-band frames.

\section{Analysis \& Results}

The MIPS observations of our two clusters yield 986 sources in the
Cl\,0024+16 field and 1071 in MS\,0451$-$03, both with 24$\mu$m flux
densities above $S_{\rm 24\mu m}>200\mu$Jy (unless otherwise stated,
quoted fluxes are always the corrected 16$''$ aperture values). In
the case of MS\,0451$-$03 the count includes sources in the archival
region in the core. We assess the false detection rate by running our algorithm on the inverse of the data, and detect $\sim 20$ sources in each frame, all with fluxes in the range 200--330$\mu$Jy, representing $\sim$2\% of the detected sample. Our analysis requires matched optical $R<24$ identifications and so we expect the false rate will be much less than 1\%. As we demonstrate below, the achieved source surface density is below 1 object per 40 beams, the classical definition of a confused map (Hogg 2001), and so our maps are not expected to be confused at this depth.

We next use the deep optical and near-infrared
imaging of these fields to investigate the photometric
properties of mid-infrared sources. The influence of false detections described above is minimised in the main part of our analysis, since we require the mid-infrared sources to have optical and near-infrared counterparts.

\subsection{Optical properties}
\label{sec:cmr}

Positional matching of multi-wavelength data sets has been a long
standing problem in astronomy, and can be particularly troublesome when
there is a large disparity in the resolution and sampling in two
datasets. We find that a simple nearest-neighbor match is not adequate
to pair mid-infrared sources with optical counterparts -- the method
can fail to match complicated interacting systems for example. Instead,
we apply the technique of de Ruiter, Arp \& Willis (1977) who use a
Bayesian estimator for the probability, $p({\rm id}\mid r)$, that a
nearby source is a true match and not a chance unrelated object.  In
Appendix~A, we briefly outline the key elements of the method, but
refer the reader to de Ruiter, Arp \& Willis (1977) for a thorough
derivation. The result of our matching analysis is a list of probable
optical counterparts for the 24$\mu$m sources in our catalogs.  We
identify 611 and 650 counterparts to 24$\mu$m sources which are
brighter than $R=24$ and 200\,$\mu$Jy in Cl\,0024$+$16 and
MS\,0451$-$03 respectively. In MS\,0451$-$03 this includes sources in
the archival region.

In Figure~1 we plot the $(B-R)$ versus $R$
color-magnitude diagram for each cluster, comparing the distribution
of colors for the mid-infrared sources and their apparent magnitudes with the 
optical population in these regions. The first thing to note is
the broad similarity in the distributions on the
color-magnitude plane of the 24$\mu$m sources in the two fields.
Looking at the color distributions in more detail, the median $(B-R)$
color of mid-infrared sources at $R<22$ is 1.75 and 1.80 for
Cl\,0024+16 and MS\,0451$-$03 respectively.  The 24-$\mu$m counterparts
exhibit a broad peak in $(B-R)$ color, which lies between the peaks of
red and blue galaxies (Fig.~1).  This association of 24$\mu$m sources
with galaxies having transition colors is intriguing.  It may simply
reflect the fact that these galaxies are dustier examples of the
general blue star-forming field population, where dust reddening
produces somewhat redder colors.  Or it may be evidence that some of
these galaxies are part of an evolutionary sequence connecting the blue
star-forming population, and the passive types inhabiting the
red-sequence.

In Figure~2 we plot the $(R-K)$ versus $(B-R)$ colors of the
mid-infrared sources in Cl\,0024+16 and MS\,0451$-$03 compared to the
optically-selected populations in the two fields. 
We also indicate the expected
colors of cluster members with a range of spectral types (King \& Ellis
1985). These model colors provide a rough guide to the relative level
of current to past star-formation in the galaxies and the reader should
note that the influence of dust extinction will move galaxies parallel
to this sequence, further complicating any interpretation of the
star-formation histories of these galaxies.

Looking at the panel for Cl\,0024+16 in Figure~2, we see a clear ridge
of 24-$\mu$m sources with colors comparable to those expected for
cluster galaxies.  To confirm this association, we indicate on the
plots the locations of spectroscopically confirmed 24$\mu$m cluster
members from the surveys of Czoske et al.\ (2001) and Moran et al.\
(2006), and our on-going spectroscopic follow-up of our 24$\mu$m sample
(Geach et al.\ in prep).  There are 45 confirmed 24$\mu$m members in Cl\,0024+16 and 7 in MS\,0451$-$03 and the vast majority of these lie close to the
predicted color-color locus for cluster members, and have colors
consistent with star-forming galaxies. However a few fall in the
region populated by passive, early-type cluster members, as expected
due to the likely presence of active galactic nuclei in some of these
galaxies.  Nevertheless, it is clear that the information from this
$(B-R)$--$(R-K)$ color-color plot can be used to crudely isolate likely
24-$\mu$m cluster members to statistically gauge the size of this
population in the two clusters.

The `transitional' colors of the majority of 24$\mu$m sources hinted
at in the color-histograms in Figure~1, is also reflected in the
distribution of points around the spectral classes in Figure~2.  The
colors of the bluer, probable 24$\mu$m cluster members suggest these
are moderately star-forming galaxies (matching the continuum shapes of
present-day Sbc/Scd galaxies).  However, these could be more actively
star-forming systems with additional reddening, $A_V \sim 0$--1, and
there are a small number of galaxies with very red colors which suggest
very significant reddening, $A_V\sim2$ mags. To select the majority of
infrared cluster members, we apply a photometric cut in color-color
space around the sequence of known cluster members
(Figure~2) and then account for the residual field contamination
using a similar color-selection applied to a blank field
24$\mu$m sample.  
Based on the existing spectroscopic
samples of 24$\mu$m sources in the two clusters, we estimate that
our color-selection criteria identify 39/45 and 5/7 members
in Cl\,0024+16 and MS\,0451$-$03 respectively, or an
average completeness of 85\%.  
We note that the fact that
the reddening vector runs approximately parallel to our selection
box in Figure~2 means that it should contain a significant fraction of
the more reddened sources.   

In addition, in this work we
are investigating the nature of the star-forming cluster population, so
we must be aware of contamination by infrared activity caused by AGN in
passive (and also star-forming) galaxies. Therefore we define a
sub-region in the color-space around the likely color of cluster E/S0s
(Figure~2),
and do not include the galaxies within this region in our analysis of
the obscured star-forming population.  

The 24$\mu$m galaxy population in the
MS\,0451$-$03 field does not exhibit such a pronounced ridge with
colors matching those expected for cluster members, as Cl\,0024+16.
Nevertheless, there is a weak overdensity,
which is most obvious in a red clump associated with passive,
early-type cluster members.  This clump is around 0.2\,mags redder in
$(B-R)$ and $(R-K)$ than the equivalent feature in Cl\,0024+16, as
expected from the $k$-correction for passive galaxies between $z=0.39$ and $z=0.55$. 
Again, the small number of spectroscopically confirmed 
24$\mu$m members fall close to the locus expected for the colors
of cluster members and we therefore use this color-color selection
in the subsequent sections to investigate the properties of this
population.

\begin{inlinefigure}
\vspace{5mm}
\centerline{\psfig{file=f4.ps,width=3.5in,angle=0}}
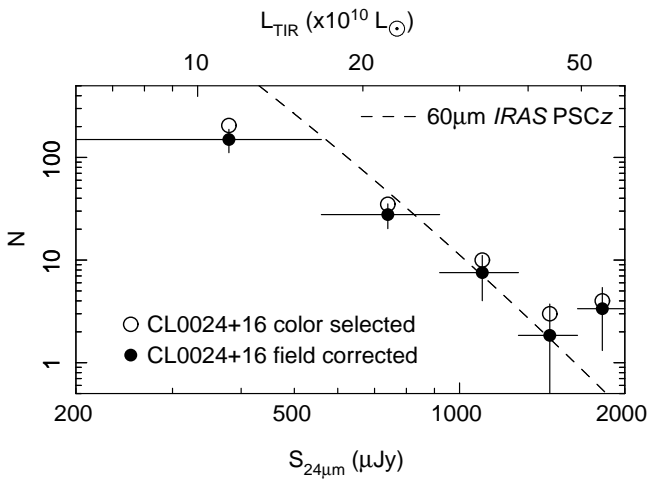
\begin{minipage}{3.5in}{{\small \sc Fig.~4.}--\small\ The infrared luminosity function for Cl\,0024+16, based on
the photometrically-selected cluster members and plotted with and
without correction for residual field contamination. The total infrared
luminosities are derived by extrapolating the monochromatic MIPS
24$\mu$m flux, assuming the local 15$\mu$m--far-infrared correlation
holds at $z\sim0.5$ (see \S3.3). The estimated number of 24$\mu$m
cluster members in MS\,0451$-$03 is insufficient to provide a
similar plot for that cluster. For comparison, we plot the luminosity
function of the 60$\mu$m {\it IRAS} Point Source Catalog Redshift 
(PSC{\it z}) survey from Takeuchi et al.\ (2003), with an arbitrary
normalisation.  The cluster luminosity function is similar in shape with that
of3 local field galaxies, given the large uncertainties.}\end{minipage}
\setcounter{figure}{4}
\end{inlinefigure}

\subsection{Counts of mid-infrared sources}

To quantify the number of 24$\mu$m sources associated with cluster
members we first compare the 24$\mu$m number counts for each cluster to
counts from the literature for large area 24-$\mu$m field surveys (e.g. Marleau et
al.\ 2004; Papovich et al.\ 2004; Chary et al.\ 2004).  The counts in
the literature all agree well in the flux density regime covered by our
data and we use the model presented in Marleau et al.\ (2004) to
estimate the number of field 24$\mu$m galaxies expected with $S_{\rm
24\mu m}=200$--1000$\mu$Jy in our fields: $dN/dS =
(416\pm3)S^{-2.9\pm0.1}$, which is valid for the range 200--900$\mu$Jy,
we extrapolate the bright end to 1000$\mu$Jy. For this comparison we
also correct our counts for completeness in the range 200--300$\mu$Jy
(our simulations show that our observations are $\sim$100\% complete at 300$\mu$Jy).

The predicted numbers of 24\,$\mu$m sources to our flux limits over the
Cl\,0024+16 and MS\,0451$-$03 fields are 1280$\pm$160 and 939$\pm$120
respectively, down to 200$\mu$Jy.  These should be compared to the
observed numbers sources of 1032 and 935, respectively.  Based on this
comparison we can see that the cluster populations do not represent a
significant excess over the large field population across the full
survey area and down to the flux limits of our survey.  This is not
particularly surprising as the survey fields are large and our
sensitivity limit is sufficient to detect sources in a large volume out
to high-redshifts.  Instead, if we restrict the comparison to the central
regions of the clusters, where the contrast of the cluster over the
background might be largest, we see that at projected radii of
$\ls2$\,Mpc there is a slight excess of mid-infrared sources compared
to the field counts in Cl\,0024+16: $134\pm12$ observed sources
compared to $115\pm14$ predicted for the field, suggesting a modest
cluster population. While in MS\,0451$-$03, including the archival data
for the core, and we find that within 2\,Mpc there are $104\pm10$
sources compared to $108\pm13$ predicted by the model -- showing no detectable
overdensity in this cluster even in the central regions.

The lack of a detectable excess in the raw counts of 24$\mu$m sources
in the cluster fields does not rule out a significant cluster
population given the large uncertainties in the contribution from field
sources.  To provide a more sensitive test we can exploit the
optical-near-infrared photometry to remove the bulk of the background
field contamination and so reduce the uncertainty in the field
correction.  In Figure~3 we plot the cumulative number of mid-infrared
galaxies as a function of clustocentric radius for those galaxies with
optical-near-infrared counterparts whose colors lie close to the
ridge-lines of cluster members seen in Figure~2 (restricted to those
with $R < 24$).  These are corrected for residual field contamination
by using the optical/near-infrared photometry of mid-infrared sources
from the SWIRE ELAIS N1 survey (Lonsdale et al.\ 2003, 2004; Surace et
al.\ 2004). To achieve this we translate our $(B-R)$--$(R-K)$
selection box in Figure~2 to the equivalent region in $(r-z)$--$(g-r)$
space for the SWIRE sample using model spectral energy
distributions (Fig.~2) to transform between the various colors.
Extrapolating the number counts of these color-selected sources to our 200$\mu$Jy limit (by fitting a power-law, $dN/dS
\propto S^{-\alpha}$, to the counts) yields an average surface density
of mid-infrared field sources with $R<24$ of $0.16\pm 0.02$ arcmin$^{-2}$
and we correct our counts using this.  We also make use of the very
large SWIRE survey area to estimate the fluctuations in the field
correction in the area of each of our radial annuli in Figure~3, by
determining the fluctuation in the field counts as a function of survey
area.  This variance is included in the uncertainty in the field
correction used in Figure~3.

By selecting 24$\mu$m sources with optical-near-infrared colors similar
to those expected for cluster members, we have substantially reduced
the field contamination in our sample.  Figure~3 demonstrates a
significant 24$\mu$m cluster population in Cl\,0024+16, with 155$\pm$18
sources across our $25'\times 25'$ field (where the uncertainty
includes the effects of clustering on the field correction).  As can be
seen the excess population is distributed across the whole field, with
the numbers continuing to increase to $\sim 5$\,Mpc where the coverage
of our data begins to decline. In contrast, in MS\,0451$-$03 we find a
much weaker excess, 28$\pm$17, which is only marginally significant.
Nevertheless, we note that our existing spectroscopic survey of this
cluster (Fig.~2) confirms that there are 24-$\mu$m-detected cluster
members, but the population appears much less numerous than in
Cl\,0024+16.  In both cases, the estimates of the total cluster
populations could be increased by $\sim 18$\% to account for the
incompleteness of our color-selection, although to be conservative we
choose not to apply this correction.

While it is clear that there are populations of 24$\mu$m cluster
members, we also need to consider lensing of background sources by
these massive clusters, which may affect the predicted numbers of
cluster members. Lensing has the effect of a boosting of flux and
dilating the projected area.  To demonstrate this we model the clusters
as simple isothermal spheres. In this case a background source will
suffer an apparent deflection in radial direction by $\alpha \simeq
20''\sigma_{1000}^2$, where $\sigma_{1000}$ is the cluster velocity
dispersion in units of 1000\,km\,s$^{-1}$. We adopt $\sigma =
1150$\,km\,s$^{-1}$ for Cl\,0024+16; $\sigma = 1290$\,km\,s$^{-1}$ for
MS\,0451$-$03 (Trentham et al.\ 1998).  At a radius $r$, the
source-plane area of an annular bin of width d$r$ will be
$2\pi(r-\alpha){\rm d}r$, and the flux of a lensed source will be
boosted by a factor $r/(r-\alpha)$. We integrate the lensing prediction for our color-selected sample
(taking $\sigma=1200$\,km\,s$^{-1}$ -- though the actual values may be
lower, so this should be a conservative estimate) as a function of
radius, finding that it could contribute $\ls 20$ sources to our samples, hence in Cl\,0024+16 the excess signal we see
is greater than can be explained by lensing.  The situation with
MS\,0451$-$03 is more problematic, although again we note that the
spectroscopic identification of a handful of 24$\mu$m cluster
members confirms that, as with Cl\,0024+16, there is a 
population of mid-infrared sources associated with the cluster.

\begin{inlinefigure}
\centerline{\psfig{file=f5.ps,width=3.5in,angle=0}}
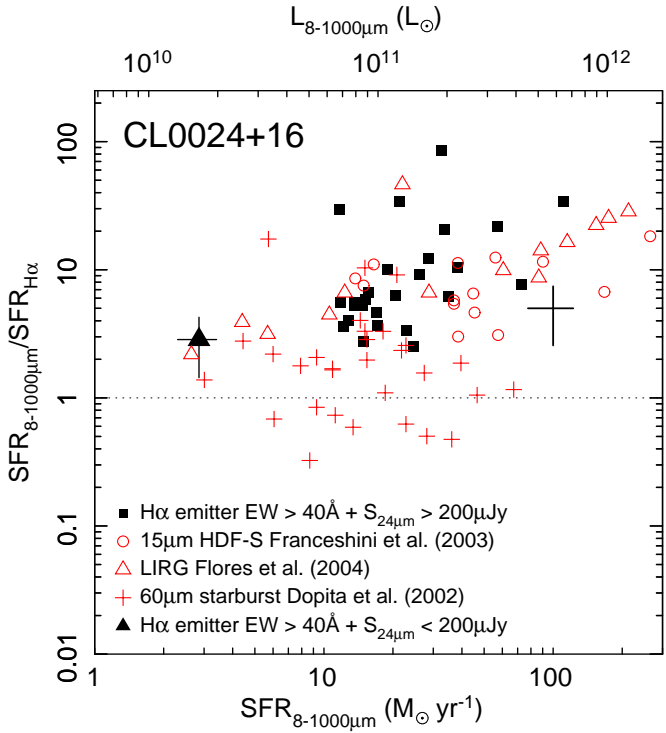
\begin{minipage}{3.5in}{{\small \sc Fig.~5.}--\small\ A comparison of the star-formation rates derived from H$\alpha$
(Kodama et al.\ 2004) and total-infrared luminosities for individual
H$\alpha$-detected cluster members in Cl\,0024+16.  The ratio
SFR[IR]/SFR[H$\alpha$] provides an estimate of the level of obscuration
of the starforming regions within these galaxies. We see a typical
offset of a factor of $5\times$ between the infrared and H$\alpha$
estimates (the latter have not been corrected for extinction), although
there is a wide-scatter.  We also plot data from the literature of
similar studies of infrared-luminous field galaxies from Flores et al.\
(2004), Franceshini et al.\ (2003) and Doptia et al.\ (2002) corrected
to our cosmology.  It appears that the level of obscuration of
star-forming infrared cluster galaxies (in Cl\,0024+16) is similar to
that of mid-infrared selected field galaxies at this epoch.  We do not plot
individual error bars for clarity, but show a representative error bar. Finally, to probe obscuration from faint mid-infrared sources, we stack the 24$\mu$m data for H$\alpha$ emitters which are not detected in the MIPS image yielding a 5$\sigma$ statistical detection. This confirms the general trend seen above our nominal luminosity limit that there appears to be an increase in the level of obscuration (SFR[IR]/SFR[H$\alpha$]) with infrared luminosity.
}\end{minipage}
\setcounter{figure}{5}
\end{inlinefigure}

\subsection{Star formation}
\label{sfrs}

To calculate the star-formation rates (SFRs) for our mid-infrared
detected galaxies, we begin by estimating the total infrared luminosity
over the restframe wavelength range 8--1000 $\mu$m. Since we only have MIPS 24$\mu$m
photometry, we have to determine the correction factor required to
convert our 24$\mu$m luminosities to total infrared luminosities. To
achieve this we assume the correlation between 15$\mu$m (corresponding
to observed $\sim$\,24\,$\mu$m at $z\sim0.5$) and total-infrared
luminosity seen for local infrared galaxies (Chary \& Elbaz 2001)
continues to hold at $z\sim0.5$. We then follow a similar technique to
that of Bell et al.\ (2005) and use the spectral energy distribution
(SED) templates from Dale \& Helou (2002) to estimate the ratio of the
observed-frame 24-$\mu$m luminosity to the 8--1000$\mu$m luminosity.
We calculate the ratio of total-infrared to observed 24-$\mu$m
luminosities for each SED template, taking the mean value of all the
ratios as our correction factor, and the range from the most active to
the most quiescent SEDs as a conservative estimate of the systematic
uncertainty. The correction factors are $16\pm2$ and $16\pm3$ for
Cl\,0024+16 and MS\,0451$-$03 respectively. We compare our conversion factors to the relation found by Chary \& Elbaz (2001) by calculating the mid-infrared to total infrared ratio with the {\it ISOCAM LW3} (15$\mu$m) filter for the Dale \& Helou templates at $z=0.1$. The Chary \& Elbaz conversion for local LIRGs corresponds to $L_{IR} \sim 11^{+6}_{-4}\times L_{\rm 15\mu m}^{0.998}$, while our estimate gives a prefactor of $7\pm 3$ over the full range of LIRG templates. Thus these two calibrations are consistent, although the reader should note our estimates are roughly 40\% fainter than would be given by Chary \& Elbaz (2001).

Our 5\,$\sigma$ 24-$\mu$m
flux limit of 200$\mu$Jy corresponds to average total infrared
luminosities of $6\times10^{10}$\,L$_\odot$ and
$12\times10^{10}$\,L$_\odot$ in Cl\,0024+16 and MS\,0451$-$03
respectively, hence the bulk of the populations we are detecting are
Luminous Infrared galaxies (LIRGs) with L$_{IR}\geq
10^{11}$\,L$_\odot$.  These luminosities translate into SFRs of
$\sim10$\,M$_\odot$\,yr$^{-1}$ and $\sim20 $\,M$_\odot$\,yr$^{-1}$
assuming the far-infrared star-formation calibration given by Kennicutt
(1998) with a Salpeter IMF and a mass range of 0.1--100\,M$_\odot$.

In Figure~4 we plot the mid-infrared luminosity function for the
Cl\,0024+16, based on the color-selected sample out to a radius of
5\,Mpc.  To field correct the luminosity
function in Cl\,0024+16, we integrate the extrapolated power-law fit to
the color-selected 24$\mu$m sources in the SWIRE ELAIS N1 field, from
\S3.2. We use the scatter in the normalisation of $dN/dS$ in
independent $25'\times 25'$ regions within the SWIRE field, combined
with the counting errors in each bin, to give the uncertainty in the
field-correction.  This plot demonstrates that the field-corrected
luminosity function in Cl\,0024+16 is similar in form to that seen for
60$\mu$m-selected populations at low-redshift (Takeuchi et al.\ 2003).

While there are too few sources in MS\,0451$-$03 to allow us to make
an equivalent plot to Figure~4 for that cluster, we can use the
Cl\,0024+16
luminosity function to estimate how many sources we would have detected
if we place the Cl\,0024+16 population at $z=0.55$.  Applying a
factor to scale for the effective areas of our surveys in
Cl\,0024+16 and MS\,0451$-$03, we estimate that we would
detect $69\pm 21$ galaxies above the luminosity limit,
$12\times 10^{10}$\,L$_\odot$, of the MS\,0451$-$03 survey.
This compares to our estimate of $28 \pm 17$ from our MS\,0451$-$03
catalog, indicating a roughly $2$--$3\times$ difference in the
numbers of mid-infrared sources in the two clusters. This suggests that, although the dearth of mid-infrared sources in MS\,0451$-$03 may be partly due to sensitivity limitations, there may be a real difference in the mid-infrared populations in the two clusters.

\subsubsection{Mid-infrared versus H$\alpha$ tracers of activity}

We can also compare the properties of our mid-infrared selected sample
in Cl\,0024+16 with the H$\alpha$ survey of this cluster by Kodama et
al.\ (2004).  Kodama et al.\ obtained a deep image of the cluster in a
narrow-band filter centered on redshifted H$\alpha$.  Galaxies with
excess emission in this narrow-band filter were identified as cluster
members with H$\alpha$ emission.  They estimate their sensitivity limit
corresponds to a star formation rate of $\sim 1.5$\,M$_\odot$\,yr$^{-1}$
-- significantly below the estimated SFR limit of our mid-infrared
survey, $\sim 10$\,M$_\odot$\,yr$^{-1}$.  

To start with, we compare in Figure~3 the cumulative radial profile of
the mid-infrared sources with the H$\alpha$-emitting galaxies from
Kodama et al.\ (2004).  As can be seen the two tracers show very
similar trends, with a steep rise from 1\,Mpc out to $\sim $5\,Mpc, 
beyond which our coverage declines. The reader
should also note that the lack of mid-infrared coverage for the core of the
cluster will contribute to the steeper drop of the 24$\mu$m counts in
the innermost bin in Figure~3.  Surprisingly, Figure~3 also shows
there are similar numbers of H$\alpha$ emitters and 24$\mu$m-detected
members, even though there is a significant difference in the relative
sensitivity of the two surveys. We therefore next compare the estimates
of the SFR for individual galaxies from H$\alpha$ and the mid-infrared, 
to relate these two star-formation indicators for
confirmed cluster members.
 
In Figure~5 we compare the star-formation rates for H$\alpha$
narrow-band-selected sources and their 24$\mu$m
counterparts\footnote{The optical star-formation rates are derived from
the H$\alpha$ and adjacent [N\,{\sc ii}]$\lambda\lambda$6548,6583
emission lines after Kennicutt, Tamblyn \& Condon (1994).}.  From this
figure, it can be seen that the star formation rates for the
mid-infrared-detected members, derived from the H$\alpha$ emission line
(but not corrected for extinction), appears to underestimate the SFR
from the mid-infrared by factors $\gs 5\times$. Including the
standard assumption of 1 magnitude of extinction and correcting
for a 30\% contribution to the measured fluxes from [N{\sc ii}]
(see Kodama et al.\ 2004) would reduce this offset to a factor
of $2.5\times$.  However, such a simplistic correction
would miss the fact that the discrepancy in SFRs
appears to increases with infrared luminosity, such that
the most active systems are also the most obscured.  

To test for biases due to our 24$\mu$m sensitivity limit, we stack thumbnail images from the MIPS image of Cl\,0024+16 for H$\alpha$ emitters from Kodama et al. (2004) which are not individually detected, i.e. $S_{\rm 24\mu m} < 200$\,$\mu$Jy. We find a $\sim$5-$\sigma$ detection with  $S_{\rm 24\mu m} = 57\pm11$\,$\mu$Jy. In terms of star-formation, this corresponds to $\sim$3\,M$_\odot$\,yr$^{-1}$. We plot this point on Fig.~5, with the corresponding average H$\alpha$ derived SFR for the stack, $\sim$1\,M$_\odot$\,yr$^{-1}$. This point seems to be roughly consistent with the trend seen, that obscuration (i.e. SFR[IR]/SFR[H$\alpha$]) increases with the level of infrared activity. 

We estimate that the correction required to bring the H$\alpha$ derived
SFRs in line with the infrared SFRs corresponds to a reddening of
$A_V\sim2.5$, slightly larger than the value used to explain the
broadband colors of these galaxies on the $(R-K)$--$(B-R)$
color--plane.  This reddening estimate will also be affected by the
choice of adopted SED when deriving total infrared luminosities,
although in this work we have chosen a conversion which represents
infrared galaxies of a wide range of activities, and so our estimates
are conservative. We also note that there is a factor of 40 range in
infrared to H$\alpha$ derived star formation rates within our sample
which is larger than expected from our measurement errors and indicates
that there may be a wide range of extinction amongst the individual
LIRGs -- in part probably due to the patchiness of the dust within
these galaxies.

We also compare our relation between mid-infrared- and
H$\alpha$-derived SFRs to that seen in samples of field galaxies from
the literature: a sample of LIRGs in the range $0.2<z<0.7$ from Flores
et al.\ (2004), 15$\mu$m detected {\it ISOCAM} sources from the {\it
Hubble Deep Field South} at ($0.2<z<1.5$) (Franceshini et al.\ 2003),
and {\it IRAS} 60$\mu$m warm infrared galaxies from Doptia et al.\
(2002). The typical level of SFR[IR]/SFR[H$\alpha$] is similar to
H$\alpha$/MIR detected objects in Cl\,0024+16, suggesting that these
cluster galaxies are similar in nature (at least in terms of their
mix of star-formation and obscuration) to star-forming LIRGs in the field at
similar epochs.

\subsection{Total cluster star-formation rates}

We now estimate a total star-formation rate for the clusters
using the typical SFR of the likely star-forming mid-infrared members. 
Since we lack observations in the central $\sim1$\,Mpc in
Cl\,0024+16 due to the unobserved GTO region, we also need to correct
for this missing population using the 15$\mu$m {\it ISO} observations from
Coia et al.\ (2005).

\subsubsection{Cl\,0024+16}

As shown by the radial number counts presented in Figure~3, the
24$\mu$m members are widely distributed across
the clusters out to a projected radius of $\sim$2\,Mpc (corresponding to $\sim R_{200}$, Carlberg\ et~al.\ 1997. Note that throughout, we use projected distances) and beyond.  To
compare the  mid-infrared-estimates of the total star formation rates
within representative regions of different clusters
we therefore adopt a 2\,Mpc radius for our calculations, which
roughly matches the size of the surveyed regions in some of the 
earlier small-field surveys.

We integrate over the luminosity function in Figure~4 for sources with
projected radii $<2$\,Mpc, correcting for the expected contamination
from field galaxies using the SWIRE data for the ELAIS N1 field (see \S3.2).  To
account for the missing sources in the unobserved GTO core region we
need to add the SFR in the 15$\mu$m members 
in this region found by Coia et al.\ (2005), after
converting their 15$\mu$m fluxes to star-formation rates using an
identical method to that presented in \S3.3. 

Our derived total-infrared luminosities are lower than those derived by Coia et al.\ (2005) by a factor of $\sim$2. This difference can be attributed to the different methods of conversion used. Coia et al.'s method involves extrapolating their observed 15$\mu$m photometry to 15$\mu$m in the rest-frame, using an SED fitted to their optical and mid-infrared photometry. This is then converted to a total infrared luminosity using the empirical conversion of Chary \& Elbaz (2001) described above. In contrast our observations correspond to $\sim$15$\mu$m in the rest-frame. Hence some of this offset is due to the differences in conversion from 15$\mu$m luminosity to total infrared emission discussed in \S3.3 and the remainder is due to the choice of SED by Coia et al.\ (2005), which overpredicted the observed 24$\mu$m fluxes. As we have stated earlier, our calculations should yield conservative limits on the SFR. We note that, with our total-infrared calibration, the 5$\sigma$ detection limit of the Coia et al.\ (2005) {\it ISO} observations corresponds to $\sim 3\times 10^{10}$\,L$_\odot$.

With our conversion, we find three galaxies in the
Coia et al.\ (2005) sample that exceed our luminosity limit of $6\times10^{10}$\,L$_\odot$, and so we can add their
SFR to our total within 2\,Mpc of the core.  We thus find a total
star-formation rate for the 40 sources within 2\,Mpc to be $1000\pm210 
$\,M$_\odot$\,yr$^{-1}$. The error is derived by boot-strap resampling the flux distribution of sources within 2\,Mpc, and integrating the new luminosity function.  The median SFR per galaxy in our sample within 2\,Mpc is $\sim16 $\,M$_\odot$\,yr$^{-1}$. We note that correcting
for the estimated incompleteness in our color-selection of cluster members
would increase our estimate of the total SFR by a factor 1.07, which corresponds to a change less than its quoted uncertainty.

For comparison,
we also note that
the H$\alpha$ observations of Kodama et al.\ (2004) detect 100 galaxies
within a similar region to that discussed here. They
find a cumulative SFR 
of $\sim 470$\,M$_\odot$\,yr$^{-1}$, above their approximate
sensitivity limit of 
SFR of 1.5\,M$_\odot$\,yr$^{-1}$, or about 4.7\,M$_\odot$\,yr$^{-1}$
per galaxy.  Restricting the H$\alpha$ sample to the
$>10$\,M$_\odot$\,yr$^{-1}$ limit of our mid-infrared survey,
we find that the H$\alpha$ survey would yield an integrated
SFR of $\sim 220$\,M$_\odot$\,yr$^{-1}$.  This confirms
the effect seen in Figure~4, that although H$\alpha$ and mid-infrared
surveys are detecting a similar population, the H$\alpha$ tracer severely underestimates the underlying activity in the most active sources in the cluster: these starbursting systems are producing dust at a more copious rate and therefore optically obscured. In addition, we can state that the mid-infrared survey detects the bulk of the total star-formation activity within this region, requiring only a $\sim$20\% correction for the star-formation activity in sources with $1$--$10$\,M$_\odot$\,yr$^{-1}$.

\subsubsection{MS\,0451$-$03}

We have seen that, compared to Cl\,0024+16, this cluster appears to be
deficient in mid-infrared sources (at least down to our luminosity limit). 
Nevertheless, even a small excess of mid-infrared sources in the cluster could contribute a non-negligible star-formation rate to the overall activity of the cluster. We therefore estimate the total star-formation rate within
$\sim$2\,Mpc using the small number of excess sources identified using
our $(R-K)$--$(B-R)$ color--color selection.  Applying the same
approach as used for Cl\,0024+16, we find a total star-formation rate
within a 2\,Mpc radius of the cluster center of $200\pm100
$\,M$_\odot$\,yr$^{-1}$. 
The median SFR within 2\,Mpc is $\sim 35$\,M$_\odot$\,yr$^{-1}$
per galaxy (reflecting the brighter luminosity limit in this cluster) . 

In \S3.3 we estimated the number of mid-infrared sources we would detect if MS\,0451$-$03 had the same mid-infrared luminosity function to Cl\,0024+16, taking into account the slightly different areal coverages between the clusters. Similarly, we can estimate the total star-formation rate in MS\,0451$-$03 if we reached a luminosity limit identical to Cl\,0024+16. Assuming the faint end of the luminosity function in MS\,0451$-$03 follows a similar shape to Cl\,0024+16, we estimate that the total star-formation rate down to $6\times10^{10}$\,L$_\odot$ in this cluster is $<460$\,M$_\odot$\,yr$^{-1}$. Although the small number of excess sources in MS\,0451$-$03 can lead to a relatively high star-formation rate (compared to what might be found using an optical tracer for example), there is a clear difference in the population of mid-infrared sources in the two clusters -- but what is the physical cause of this? 

\bigskip

Although
the clusters are of similar mass, and at a similar redshift, they
differ strongly in their intra-cluster environments and dynamical
status (Treu et al.\ 2003). 
Cl\,0024+16 is dynamically active (Czoske et al.\ 2001; Kodama
et al.\ 2004) and this may provide the impetus for the triggering of
star-formation via mergers and interactions of gas-rich spirals within
the cluster -- even in the apparently high-density core. The
intracluster medium (ICM) of MS\,0451$-$03 is much hotter and 
denser (by nearly an order of magnitude) than Cl\,0024+16, and this will
have an impact on the radii that processes such as starvation and
ram-pressure stripping operate (Treu
et al.\ 2003).   
In MS\,0451$-$03 the hot ICM will be much more
effective at starving
galaxies of their gas reservoirs at a larger clustocentric radii than
in Cl\,0024+16. 
The relative dearth of mid-infrared sources in
MS\,0451$-$03 might then suggest that the active regions within
the mid-infrared population are comparatively sensitive to 
these gas removal mechanisms. Unlike Cl\,0024+16, unfortunately there are no published optical SFR studies of MS\,0451$-$03 to confirm a wholesale decline in the SFR in this cluster. Nevertheless, the small excess
of mid-infrared sources in MS\,0451$-$03 show that it hosts quite
significant star-formation, so
it is unclear at this stage exactly what physical processes control the
distribution of star-forming galaxies in different types of clusters.
We will address the issue in our next paper where we study the
properties of the 24$\mu$m cluster members in more detail using the
available spectroscopy, dynamics 
and {\it HST} imaging (Geach et al., in prep.; see
also Moran et al., in prep.).

\section{Obscured star-formation in clusters out to $z\sim1$}

At the present-day, 
virialised regions of the Universe 
are dominated by passive
galaxies: ellipticals and lenticulars, with the main contribution to
the global star-formation rate density at $z=0$ coming from late-type
spirals in low-density environments. However, at higher redshifts the
high-density environments within clusters show increasing numbers of
actively star forming galaxies, possibly reflecting a similar increase
in the frequency of activity to that seen in the surrounding field.
Moreover, it is clear that there is significant hidden star formation
in clusters in this redshift range: observations by {\it ISO} and now
our new {\it Spitzer} observations reveal populations of infrared
galaxies in clusters at low and intermediate redshift, with SFRs much
higher than would be measured using optical tracers such as H$\alpha$
or [O{\sc ii}]$\lambda$3727, due to the extinction effect of dust.

Given this new information, what is the evolution of the total
star-formation rates in massive clusters -- i.e.\ how does the increase
in activity in high-density environments at high redshift behave in the
mid-infrared? To compare the total mid-infrared derived star-formation rate
in clusters from $z=0$--1 we use the total SFR within $\sim$2\,Mpc
and normalize by the best-estimate mass of the cluster (for example, see Kodama 
et al.\ 2004). For Cl\,0024+16
and MS\,0451$-$03 these are $6.1\times10^{14}M_\odot$ (Kneib et al.\
2003) and $15\times10^{14}M_\odot$ (La Roque et al. 2003) respectively, where the masses are within 2\,Mpc of the clusters' centers. 

To build up
a history of star-formation in clusters out to $z\sim1$, we compare our
results to previous {\it ISO} and {\it IRAS} studies of the clusters:
Perseus (A\,426), A\,1689, A\,370, A\,2218 and J1888.8\,CL
(Cl\,0054$-$27), summing the known cluster member's star formation
rates down to a far infrared limit $6\times10^{10}L_\odot$ -- the limit
of the Cl\,0024+16 observations. To ensure correct comparison of the
rates, if necessary we re-derive the SFRs using our employed
calibration from Kennicutt (1998), with a standard Salpeter IMF with a
mass range 0.1--100$M_\odot$. For the {\it ISO} observations, we also
re-derive the far-infrared 8--1000\,$\mu$m luminosity using the method we
present above, using the band-pass of the {\it ISO LW3} filter to
convert mid to total-infrared.  The {\it ISO} observations of these
clusters are summarized in the review by Metcalfe et al.\ (2005). Note that the previous {\it ISO} and {\it IRAS} observations have concentrated on the core regions of clusters (typically within 1\,Mpc), so we restrict our star-formation integration to within a radius of 2\,Mpc in Cl\,0024+16 and MS\,0451$-$03, and compare to estimates of the total star-formation within the equivalent radius in the other clusters. In the case of MS\,0451$-$03, we estimate an upper limit to the star-formation rate extrapolated to a luminosity limit of $6\times10^{10}$\,L$_\odot$ by using the method described in \S3.4.2, assuming the faint end of the luminosity function follows a similar shape to that in Cl\,0024+16. We estimate that the upper limit to the total star-formation rate down to $6\times 10^{10}$\,L$_\odot$ in MS\,0451$-$03 is $\ls460$\,M$_\odot$\,yr$^{-1}$. 

\begin{inlinefigure}
\vspace{5mm}
\centerline{\psfig{file=f6.ps,width=3.5in,angle=0}}
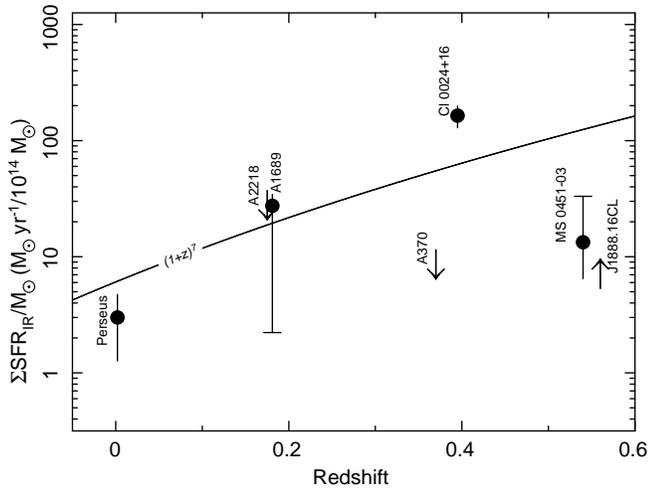
\begin{minipage}{3.5in}{{\small \sc Fig.~6.}--\small\ The variation in the mass-normalised star-formation rates in
clusters out to $z\sim0.5$. The star formation rates are from the
mid-infrared populations within $\sim$2\,Mpc and these are normalised
to the best estimate of the cluster mass, which is derived via lensing
estimates, or from the X-ray luminosities. We also plot an evolutionary
model for the counts of star-forming ULIRGs from Cowie et al.\ (2004), normalised to the mean star-formation
rate in Cl\,0024+16 and MS\,0451$-$03 (again we note that these
could be increased by $\sim$\,18\% to account for
incompleteness in our color-selection  of cluster members). The MS\,0451$-$03 point has an upper limit which corresponds to the extrapolated estimate for the total star-formation down to the luminosity limit in Cl\,0024+16 ($6\times 10^{10}$\,L$_\odot$), while the data point shows the cluster's summed SFR down to the limit of our data ($12\times 10^{10}$\,L$_\odot$). There is evidence
for an increasing rate of activity in more distant clusters, as traced
through their mid-infrared populations. However, there is also
clear evidence for a wide variation in activity in even massive
clusters at a similar epoch.  This suggests that the mid-infrared
populations are sensitive tracers of environmental changes within
the clusters.  The errors on the total
star-formation rates for Cl\,0024+16 and MS\,0451$-$03 are derived
from boot-strap resampling of the mid-infrared distribution, whereas for Perseus and A\,1689 the errors are from counting statistics. Note that there are systematic uncertainties in all of the estimates
depending on the specific choice of SED in the conversion from
mid-infrared luminosity to star formation rate (the plotted uncertainties also do not take into
account of the errors in the cluster mass estimates). Upper limits are equivalent star-formation rates for one detected at our luminosity limit of 6$\times$10$^{10}$\ L$_\odot$, (and extrapolated to account for the coverage difference in the survey) in those clusters where no sources were detected above this limit.}\end{minipage}
\end{inlinefigure}

At low redshifts, we use the nearby rich cluster Perseus,
observed with {\it IRAS} (Meusinger et al.\ 2000). We sum over the known
cluster members with L$_{\rm FIR} > 6\times10^{10}$\,L$_\odot$, and convert
their luminosities to SFRs, giving a lower limit to the total star-formation rate in
the cluster of $>22$\,M$_\odot$\,yr$^{-1}$, over an area equivalent to 10\,degrees$^2$, which in terms of physical coverage is similar to the 2\,Mpc radius used in Cl\,0024+16 and MS\,0451$-$03.  We use the mass estimate of
Ettori, De Grandi \& Molendi (2002) of $3.1\times10^{14}$\,M$_\odot$ and extrapolate to 2\,Mpc to determine the normalised star-formation rate per mass to be $\sim 3
$\,M$_\odot$\,yr$^{-1}$/$10^{14}$\,M$_\odot$.

Fadda et al (2000) and Duc et al.\ (2002) observed A\,1689 ($z=0.18$) at 15\,$\mu$m with {\it
ISO} detecting 16 cluster members within 0.5\,Mpc of the
core. With the mid-infrared to total-infrared conversion used in this work, we find two galaxies (detected with the {\it LW3} 15$\mu$m filter) above our luminosity limit. In order to estimate the star-formation expected out to a radius of 2\,Mpc we assume a radial profile for the mid-infrared population similar to that found for Cl\,0024+16 and extrapolate the star-formation rate from within a radius of 0.5 to 2\,Mpc. The resulting star-formation rate is $\sim$280\,M$_\odot$\,yr$^{-1}$ within 2\,Mpc of the core.  We use the mass from King et al. (2002), correcting to 2\,Mpc, yielding a mass-normalized value of
$\sim$30\,M$_\odot$\,yr$^{-1}$/$10^{14} M_\odot$. A\,2218 is another rich
cluster at a similar redshift to A\,1689, with $z=0.175$, however the
mid-infrared activity in this structure is much lower than in
A\,1689. The 15$\mu$m observations of Biviano et al.\ (2004) found nine
members within a similar radius ($\ls0.4$\,Mpc), but only one of these
corresponds to a star-forming galaxy, and the median infrared
luminosity is only $6\times10^8$\,L$_\odot$, or just 0.1\,M$_\odot$\,yr$^{-1}$. A\,370 is at a similar redshift to
Cl\,0024+16, at $z=0.37$, but also appears deficient in LIRGs. Only one
cluster member was detected at 15$\mu$m. We plot these points as upper
limits using our luminosity limit, again extrapolating to account for radial coverage out to 2\,Mpc using the shape of the radial profile of mid-infrared sources in Cl\,0024+16. We use the cluster masses from Pratt et al. (2005)  and Girardi \& Mezzetti  (2001)  for A\,2218 and A\,370 respectively, extrapolating to find the mass within 2\,Mpc. The mass-normalised upper-limits for A\,2218 is $30$\,M$_\odot$\,yr$^{-1}$/$10^{14}$\,M$_\odot$ and for A\,370 is $9$\,M$_\odot$\,yr$^{-1}$/$10^{14}$\,M$_\odot$.

J\,1888.16 (Cl\,0054-27) is at $z=0.56$, and was observed by Duc et
al.\ (2004) using {\it ISOCAM} at 15$\mu$m. Six mid-infrared sources
were detected which are confirmed cluster members, whilst a further two
have redshifts suggesting they are members of infalling groups at a
slightly higher redshift. Using the calibration of Chary \& Elbaz
(2001), Duc et al.\ find that all of their detected members are LIRGs, with
inferred L$_{\rm FIR}>1.3\times10^{11}$\,L$_\odot$, and with individual
SFRs in the range 20--120\,M$_\odot$\,yr$^{-1}$. We re-calibrate the SFR
using the method set out in \S 3.3, and find a conservative lower limit
to the total cluster star-formation rate to be $>70
$\,M$_\odot$\,yr$^{-1}$. The mass of the cluster is determined by Girardi
\& Mezzetti (2001), and we extrapolate to 2\,Mpc, giving a total star
formation rate $>7 $\,M$_\odot$\,yr$^{-1}$/$10^{14}$\,M$_\odot$.

We present the results in Figure~6, in which we plot the sum of the
SFRs within $\sim$2\,Mpc of the clusters' cores, normalized using the 
estimated total mass of the cluster (either
based on lensing or X-ray models).  There is strong evolution in the
star-formation rates in the clusters out to high redshifts, but it
is important to note that several clusters seem to have very low-levels
of activity, below our luminosity limits. For example Cl\,0024+16 appears
significantly more active than A\,370, which is at an almost
identical redshift. This might be point to differing environmental
influences between clusters being the dominant influence on the
star-formation histories of in-falling galaxies. Nevertheless, assuming
that we have only selected those clusters with significant activity in
Figure~6, there still appears to be strong evolution in the sample out
to $z\sim0.5$. This observation could simply reflect the increase
in number of star-forming galaxies seen in clusters and the field out
to this redshift -- these obscured systems representing the high
luminosity tail of the general `blue' population. The idea that the
cluster (obscured) star-formation histories mimic that of the field is
supported by the rough consistency with the $(1+z)^7$ trend found by Cowie
et al.\ (2004) for the number of star-forming ULIRG radio sources out to
$z\sim1.5$. Kodama et al. (2004) present a similar analysis for the evolution of the H$\alpha$-derived star-formation in clusters out to $z\sim1$. They find relatively strong evolution in the total-SFRs in clusters over this range, $\sim(1+z)^4$, but as in this study, there is considerable scatter in the total star-formations rates between clusters, even after mass-normalisation. This hints that, although there might be a nominal rise in the level of star-formation in clusters out to $z\sim1$, this is mitigated by the fact that individual cluster environments have a strong influence on the star-formation histories of their constituent galaxies. This scatter is due to the complexity of processes operating solely in the dense environments. A detailed study of the sources in Cl\,0024+16 and MS\,0451$-$03 may elucidate this issue.

\section{Conclusions \& Summary}

We have used the MIPS instrument on {\it Spitzer} to survey the
24$\mu$m populations of two optically rich clusters at $z\sim0.5$:
Cl\,0024+16 and MS\,0451$-$03. The samples are $\sim80$\% complete at
$200\mu$Jy, corresponding to total (8--1000$\mu$m) infrared
luminosities of $6\times10^{10}$\,L$_\odot$ and
$12\times10^{10}$\,L$_\odot$ at $z=0.39$ and
$z=0.55$ respectively, equivalent to  minimum SFRs of
$\sim10$\,M$_\odot$\,yr$^{-1}$ and $\sim20$\,M$_\odot$\,yr$^{-1}$. 
We detect a total of 986 and 1018 mid-infrared
sources above this flux limit across $\sim25'\times25'$ fields in
Cl\,0024+16 and MS\,0451$-$03.
Our observations probe from around the cores out to the
turn-around radius at $\sim5$\,Mpc where the clusters merge into the
field. In MS\,0451$-$03, we also analyse archival MIPS observations of
the central $\sim5'\times5'$ of the cluster, which our observations
had to avoid. Similarly in Cl\,0024+16 we make use of an {\it ISO}
15$\mu$m survey from Coia et al.\ (2005) in the central region to build up a
picture of the distribution of mid-infrared sources over the complete 
range of cluster environments.

We exploit 
optical-near-infrared colors for the mid-infrared sources to
reduce the background field
contamination. We 
find a statistical excess of mid-infrared sources (within $\sim 5$\,Mpc of the cluster core) at $S_{\rm 24\mu
m} > 200\mu$Jy associated with Cl\,0024+16: 155$\pm$18. 
In contrast MS\,0451$-$03 has a less significant population of
mid-infrared sources, $28\pm17$, although we note that
there are a small number of confirmed 24$\mu$m members in MS\,0451$-$03
in our on-going spectroscopic survey of this cluster.
 
Using our deep optical and near-infrared imaging of both
clusters we show that the 24$\mu$m sources in Cl\,0024+16 are mostly
associated with star-forming galaxies, with typically blue $(B-R)$
colors, but which can be dust reddened by up to $A_V\sim2.5$ mags. We
also compare the infrared star-formation rate to that derived from an
optical narrow-band H$\alpha$ survey of this cluster from Kodama et
al.\ (2004). Typically the H$\alpha$-derived rates underestimate the
extinction-free infrared rate by $\gs5\times$, suggesting significant
obscuration of the activity in this cluster. We find that the level of
obscuration for these individual cluster galaxies is comparable to that
found for LIRGs in the field at similar epochs. This suggests that
starbursts in clusters are similar (at least in terms of extinction) as
those triggered in low-density environments. However the variation in
the 24$\mu$m populations of Cl\,0024+16 and MS\,0451$-$03 suggests that
the range of triggering and suppression mechanisms in clusters is
complex.

We estimate that the total star-formation rate (derived from the
infrared) in the central region of Cl\,0024+16 ($\ls R_{200}$) is
$1000\pm210$\,M$_\odot$\,yr$^{-1}$. MS\,0451$-$03 is much poorer in
mid-infrared sources, and we derive a total star-formation rate
estimate by summing over the small excess of objects, giving a total
star-formation rate of $200\pm100$\,M$_\odot$\,yr$^{-1}$ within the
same physical radius.  We note however that our mid-infrared survey can miss some star-formation if a substantial number of
lower-luminosity galaxies also exist in these clusters
(as shown by the H$\alpha$ survey of Kodama et al.\ 2004). However, we
show that the majority of the activity is dominated by these
dusty-starbursts.

Finally, we look at the evolution of the specific star formation rate
per cluster with redshift from $z\sim 0$--0.5, using our new
estimates for the total star-formation rates in Cl\,0024+16 and
MS\,0451$-$03.  We compare these to estimates for lower and comparable
redshift clusters studied with {\it ISO} and {\it IRAS}. We find that
the high redshift clusters tend to have larger total star-formation
rates compared to the more quiescent low-redshift ones, with an
evolution similar to that of field LIRGs. However there is considerable
scatter in this relation, and the evolution may only apply to the most
active clusters. Although it is still unclear exactly what processes
govern the star-formation histories of rich clusters, this study has
shown that rich environments can sustain significant amounts of hidden
star-formation, and that this seems to increase at least out to
$z\sim0.5$. This hidden activity may have a profound influence on the life-cycle of galaxies in high-density regions and the formation of the passive
galaxy populations, ellipticals and S0s, which inhabit these
environments at the present-day. We will investigate these issues in
more detail in our next paper where we bring together spectroscopic and
morphological information on the 24$\mu$m population in these fields. We are also extending our survey with new panoramic observations of distant clusters with {\it Spitzer} Cycle 3 GO time using both MIPS and the IRS.

\acknowledgements

We thank an anonymous referee, whose comments greatly improved the clarity of this paper, we also thank Bianca Poggianti and Mark Swinbank for useful comments. This study is based on observations made with the {\it Spitzer Space
Telescope}, which is operated by the Jet Propulsion Laboratory,
California Institute of Technology under a contract with NASA. This
work also made use of the {\it SST} Archive, which is operated by the
{\it Spitzer} Science Center. J.E.G. is supported by a UK Particle Physics and Astronomy Research
Council studentship; I.R.S. and G.P.S. acknowledge support from the
Royal Society.

\appendix

\section{Matching of mid-infrared and optical catalogs}

\noindent Here we provide a brief description of the technique which we have
adopted for identifying the optical counterparts to the mid-infrared
sources in our catalog.  We have chosen to apply the technique of de
Ruitter, Willis \& Arp (1977) who use a Bayesian estimator for the
probability, $p({\rm id}\mid r)$, to identify whether an optical source
in close proximity to a mid-infrared source is likely to be a true
match, rather than a chance unrelated object.

First we parameterize the radial distances of sources in terms of a
dimensionless variable:
\begin{equation}
r = \left( \frac{\Delta\alpha}{\sigma_\alpha}^2 + \frac{\Delta\delta}{\sigma_\delta}^2 \right)^{1/2}
\end{equation}
where $\Delta\alpha(\delta)$ are the positional offsets of the sources
in R.A. and Dec. (in the sense mid-infrared $-$ optical), and
$\sigma_{\alpha(\delta)}$ are the uncertainties in the positions (we
assume that the optical uncertainties are negligible). A likelihood
ratio $LR$ can be constructed representing the probability of finding a
genuine match (id) compared to the probability of finding a confusing
(c) source within a radius $r$:
\begin{equation}
\label{lr}
LR(r) = \frac{p(r \mid {\rm id})}{p(r \mid {\rm c})} =  \frac{1}{2\lambda}\exp\left[\frac{r^2}{2}(2\lambda - 1) \right]
\end{equation}
where $\lambda = \pi\sigma_\alpha\sigma_\delta\Sigma$, with $\Sigma$
representing the surface density of optical sources down to some
limit. The physical interpretation of equation \ref{lr} is that we are
searching for a trade-off between finding a confusing background
source, the distribution of which is governed by Poisson statistics;
and the probability that the genuine match is located within ${\rm d}r$
of the reference source -- this is described by the Rayleigh
distribution.

In our case $\lambda \sim 20\Sigma$, and since we know that a cluster
exists and the surface density of sources will not be uniform across
the frame, we calculate the surface density of $R$-band sources
brighter than the nearest-neighbour -- $\Sigma(m > m_{\rm NN})$ -- in
annular bins of width $1'$ centered on the cluster at the radial
distance of each mid-infrared source. The final part of the calculation
requires us to calculate $p({\rm id}\mid r)$ -- {\it having found} a
source at some radius $r$, what is the probability it is the genuine
match?  Bayes' theorem provides a way to estimate this value, but first
we require an {\it a priori} estimate -- namely what fraction $\theta$
of mid-infrared sources have optical counterparts? To estimate $\theta$
we count the number of sources with $LR > 1$, over the frame, denoting
these `matches' and compare to the number with $LR < 1$. $\theta$ is
insensitive to the choice of likelihood threshold, and is approximately
0.77 for Cl\,0024+16 and MS\,0451$-$03.  We can then find:
\begin{equation}p({\rm id}\mid r) =\frac{}{} =\frac{ X\,LR(r)}{X\,LR(r) +1}\end{equation}
where $X = \theta/1-\theta$. We define a positive match of sources at
some cut-off value $L$ of the likelihood ratio, which we optimize to
provide the best completeness $C$ and reliability $R$ of the sample:
\begin{equation}
C = 1 - \sum_{LR_i < L} p_i({\rm id \mid r}) / N_{\rm id}
\end{equation}
\begin{equation}
R = 1 - \sum_{LR_i > L} p_i({\rm c \mid r}) / N_{\rm id}
\end{equation} 
with $N_{\rm id} = \sum p({\rm id \mid r})$. We choose a value of
$p({\rm id \mid r}) = 0.82$ for positive matches, corresponding to $R =
C \sim 97$~\% in both clusters. To improve the matching algorithm, we
perform an identical calculation for the next-nearest neighbour to the
mid-infrared sources. Thus for each MIPS source we have a
matching probability for both its nearest and next-nearest optical
neighbours. This allows us to flag possible mergers (where both
probabilities exceed our match threshold). If $p_{\rm NNN} > p_{\rm
NN}$ then we chose the next-nearest neighbour as the genuine match.

\end{document}